\documentclass[journal=nalefd]{achemso}
\usepackage{graphicx,color}
\usepackage{amssymb,amsmath}
\usepackage{amsmath,float}

\usepackage{soul}
\usepackage{color}
\newcommand{\revision}[1]{\textcolor{black}{#1}}

\author{Xue-Ya Mi}
\affiliation[School of Physics, HUST]{School of Physics and Wuhan National High Magnetic Field Center, Huazhong University of Science and Technology, Wuhan 430074, P. R. China}
\altaffiliation{These authors contributed equally to the work}
\author{Xiaoxiang Yu}
\affiliation[\revision{NICE, School of Energy and Power Engineering, HUST}]{\revision{Nano Interface Center for Energy (NICE),} School of Energy and Power Engineering, Huazhong University of Science and Technology, Wuhan 430074, P. R. China}
\altaffiliation{These authors contributed equally to the work}
\author{Kai-Lun Yao}
\affiliation[School of Physics, HUST]{School of Physics and Wuhan National High Magnetic Field Center, Huazhong University of Science and Technology, Wuhan 430074, P. R. China}
\author{Xiaoming Huang}
\affiliation{School of Energy and Power Engineering, Huazhong University of Science and Technology, Wuhan 430074, P. R. China}
\author{Nuo Yang}
\email{nuo@hust.edu.cn}
\affiliation[\revision{NICE, School of Energy and Power Engineering, HUST}]{\revision{Nano Interface Center for Energy (NICE),} School of Energy and Power Engineering, Huazhong University of Science and Technology, Wuhan 430074, P. R. China}
\alsoaffiliation[State Key Laboratory of Coal Combustion, HUST]{State Key Laboratory of Coal Combustion, Huazhong University of Science and Technology, Wuhan 430074, China}
 \author{Jing-Tao L\"u}
\email{jtlu@hust.edu.cn}
\affiliation[School of Physics, HUST]{School of Physics and Wuhan National High Magnetic Field Center, Huazhong University of Science and Technology, Wuhan 430074, P. R. China}
\alsoaffiliation[BIR]{Beida Information Research, Tianjin 300457, China}

\title{Enhancing thermoelectric figure-of-merit by low-dimensional electrical transport in phonon-glass crystals}

\begin{document}
\begin{abstract}
Low-dimensional electronic and glassy phononic transport are two important
ingredients of highly-efficient thermoelectric material, from which two
branches of the thermoelectric research emerge. One focuses on controlling
electronic transport in the low dimension, while the other on multiscale phonon
engineering in the bulk. Recent work has benefited much from combining these
two approaches, e.g., phonon engineering in low-dimensional materials. Here,
 we propose to employ the low-dimensional electronic structure in bulk
phonon-glass crystal as an alternative way to increase the thermoelectric
efficiency.  Through first-principles electronic structure calculation and
classical molecular dynamics simulation, we show that the $\pi$-$\pi$ stacking
Bis-Dithienothiophene molecular crystal is a natural candidate for such
an approach. This is determined by the nature of its chemical bonding. Without
any optimization of the material parameter, we obtain a maximum
room-temperature figure of merit, $ZT$, of $1.48$ at optimal doping, thus validating our idea.
\\
Keywords: Thermoelectric Effect, Molecular Crystals, Electronic Structure Calculation, Molecular Dynamics Simulation
\end{abstract}
	
\maketitle
\section{Introduction}
The thermoelectric efficiency of a material is characterized by the
dimensionless figure of merit $ZT = \sigma S^2 T/(\kappa_{\rm e} + \kappa_{\rm
ph})$, with $S$ the Seebeck coefficient, $\sigma$ the electrical conductivity,
$\kappa_{\rm e}$ the electron thermal conductivity, $\kappa_{\rm ph}$ the
phonon thermal conductivity, and $T$ the temperature.  The optimization of $ZT$
is a highly non-trivial task, e.g., the electrical conductivity and the
electron thermal conductivity are, in many cases, related by the
Wiedemann-Franz law.

Different strategies have been proposed to increase the $ZT$
value\cite{snyder_complex_2008,gang,dresselhaus_new_2007,dubi_colloquium:_2011}.
In their seminal work, Hicks and Dresselhaus proposed to increase $S$ utilizing
the sharp change of electrical density of states (DOS) in low-dimensional
structures\cite{hicks_effect_1993,hicks_thermoelectric_1993,mahan_best_1996},
such as semiconductor quantum wells, quantum wires, quantum dots, and single
molecular devices\cite{dresselhaus_new_2007,dubi_colloquium:_2011}.  Another
approach is to search or design electron-crystal-phonon-glass materials by
phonon engineering (phononics)\cite{phononics11,WaWaLu07}.  The idea is to
reduce the phonon thermal conductivity while keeping the electrical
conductivity intact \cite{snyder_complex_2008}. Recently, combining these two
approaches, phonon engineering is used in low-dimensional structures, like silicon nanowires, to
boost their $ZT$
\cite{hochbaum_enhanced_2008,boukai_silicon_2008,markussen_surface-decorated_2009,Brovman2013,
Qian2014}. A natural question then arises:  Is it possible to do the opposite, utilizing low-dimensional electrical transport in bulk phonon-glass crystals?  To show this is
indeed possible, we need to start from bulk materials with low thermal
conductivity.

\revision{Thus far, the main focus of the thermoelectric research community is inorganic semiconductor materials,\cite{gang,Zhao2014} including Bi$_2$Te$_3$,\cite{Poudel02052008} PbTe,\cite{Biswas} SiGe,\cite{gangSiGe} SnSe,\cite{Zhao2014Nat} perovskites,\cite{Koumoto2010} and Si.\cite{Nuo2013,Yang14} Instead, we focus on organic semiconductors, especially molecular crystals, which have been intensively studied in the development of organic photovoltaic cells\cite{Parida20111625}, light-emitting diodes\cite{ThejoKalyani20122696}, and field-effect transistors.\cite{ADMA:ADMA99, dong2013} Currently, their thermoelectric properties have received more attention.\cite{Zhang2009-1,Zhang2009-2,wang_modeling_2012,chen_2012,zhang_organic_2014,casian13} Compared with inorganic materials, the advantages of organic thermoelectric materials are lightweight, flexible, cheap and easy to process over large areas.\cite{ Leclerc2003}} Due to the weak bonding between different molecules, organic molecular crystals have natural low thermal conductivity. Recent experimental progress has resulted in several orders of magnitude increase of $ZT$, from  $0.001\sim0.01$\cite{Mateeva1998,Pfeiffer1998} to
0.42\cite{Yue2012,Kim2013}.  Moreover, it is now possible to design and
modulate the transport characteristics of organic
structures\cite{zhang_organic_2014}.  Among them, $\pi$-$\pi$ stacking
molecular crystals have proved their superior electronic transport properties
along the development of organic field-effect transistors\cite{dong2013}.
Interestingly, it has been shown that, their electronic transport properties can be controlled by tuning
the stacking angle and distance\cite{giri_tuning_2011,dong2013}.

\revision{Here, we attend to Bis-Dithienothiophene molecular crystal (BDTMC) which has the interesting electrochemical and optical properties.\cite{Ozturk200511055} As shown in  Fig.~\ref{fig:stru}(a), each dithienothiophene molecular possess three fused thiophene rings and they form crystal structure through $\pi$-$\pi$ stacking. The way that small molecules stack together determines the structure and electronic property of the organic crystal. Herringbone stacking is one of the common stacking structures, whose electronic band structure is normally two-dimensional, as a result of $\pi$-$\pi$ stacking extending in two dimensions. The BDTMC favors coplanar stacking\cite{Sirringhaus1997} and has high mobility.\cite{Zhang2009-2,Sirringhaus1997} Due to the strong $\pi$-$\pi$ overlap only along one direction, it shows a quasi-one-dimensional (Q1D) band with a large band dispersion (Fig~\ref{fig:stru}(a-b)). This Q1D band structure is different from common two-dimensional band in molecular crystals and appropriate for our study on thermoelectrics. Besides, its thermoelectric transport properties are under experimental investigation.\cite{qiu2015}
}

Based on a general one-dimensional tight-binding
model validated by Density Functional Theory (DFT), we employ the
semi-classical Boltzmann transport theory to study its electronic transport
properties.  Meanwhile, the phonon thermal
conductivity is calculated using the classical molecular dynamics (MD)
simulation. We show that, Q1D electronic transport and
phonon-glass-like thermal transport are realized at the same time in bulk BDT
molecular crystal, which suggests a great thermoelectric potential for
$\pi$-$\pi$ stacking organic molecular crystals.

\section{Results and Discussion}
\subsection{Electronic structure.}
\begin{center}
\begin{figure}
\includegraphics[scale=0.6]{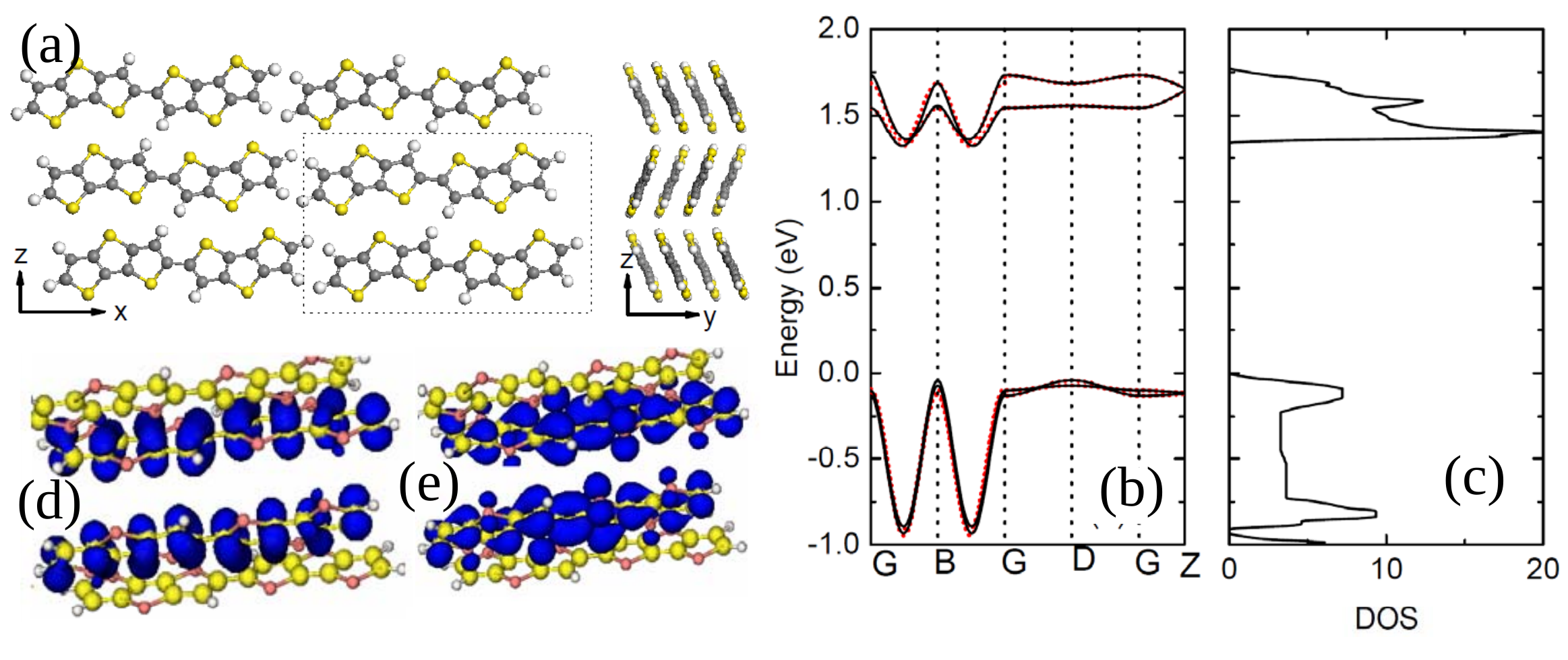}
\caption{\textbf{The lattice and DFT band structure.} (a) The relaxed structure of the BDT crystal along different
directions. The lattice vectors are: $a_1$ = (16.82, 0.087, 0.015) \r{A},
$a_2$ = (16.40, 3.85, 0.015) \r{A}, and $a_3$ = (-2.08, -0.24, 10.68) \r{A}.  The
$y$-direction is approximately perpendicular to the molecular plane.  (b)-(c)
DFT band structure and density of states (DOS) of the BDT crystal. The
energy zero is set to the top of the valence band. The reciprocal space
coordinates of high-symmetry points are G = (0, 0, 0), B = (-0.5, 0.5, 0), D = (0.5,
0.5, 0), and Z = (0, 0, 0.5), in unit of the reciprocal lattice constants.  The red
dots represent the bands fitted using one-dimensional band dispersion
$\varepsilon \sim -2t \cos(k L_0)$, with $t = -0.2, -0.2$ eV for the two valence
bands, and $-0.05, -0.08$ eV for the two conduction bands (see text for
details). (d) Charge density distribution of the two nearly degenerate HOMO
orbitals in one isolated unit cell. They are localized in each molecule, and
form the two valence bands in the lattice structure.  (e) Corresponding LUMO
orbitals forming the conduction bands.  }
\label{fig:stru}
\end{figure}
\end{center}

The lattice and electronic structure of BDTMC are calculated from Vienna Ab-initio
Simulation Package (VASP)\cite{kresse_efficiency_1996}.  The
Perdew-Burke-Ernzerhof (PBE) version of  generalized gradient approximation
(GGA) is used for the exchange-correlation
potential\cite{perdew_generalized_1996}. The DFT-D2 method of Grimme is used to
take into account the van der Waals (VDW) interactions between different
molecules\cite{grimme_accurate_2004} (details in Supporting information (SI), I(A)).  Figure~\ref{fig:stru}(a) shows the relaxed structure.
It forms a triclinic Bravais lattice, with two molecules within one  primitive
unit cell, shown in dashed lines. The conduction (CB), valence band (VB)
structure and DOS are shown in Fig.~\ref{fig:stru}(b) and (c). We observe a
strong band dispersion along $y$ direction, G-B-G, comparing to other
directions.  This indicates a large overlap of $\pi$ orbitals between molecules
along $y$ direction, due to the small angle between $y$ and the $\pi$-$\pi$
stacking ($\pi$) direction.  The band structure along $y$ is well re-produced
by a fitting using the dispersion of 1D tight-binding model (red dots). Model
details are given in SI, II.

Due to a small electronic overlap between molecules in the same unit cell, the
conduction and valence bands come in pairs, each mainly localized in one
molecule. By comparing the charge distribution of the bands with the electronic
states of the isolated molecule pair, we find that the VB and CB are simply
formed by the highest occupied molecular orbitals (HOMOs) and lowest
unoccupied molecular orbitals (LUMOs) of each molecule, respectively
[Fig.~\ref{fig:stru} (d) and (e)]. The strong anisotropy in band dispersion
suggests the formation of Q1D band structure along $y$ direction. This is
further supported by the van-Hove-like high DOS near the band edges
[Fig.~\ref{fig:stru}(c)]. For 1D system, the DOS is inversely
proportional to the hopping element $t$. Thus, the DOS at valence band edge is
around one-third of DOS at the conduction band edge. Hereafter, we focus on
the electrical properties along $y$ direction.

\revision{The intrinsic transport mechanism in organic molecular crystals has
  been under active debate. Band and hopping models have been used to
  understand the experimental results.\cite{Coropceanu2007,Pernstich2008} It is
  commonly accepted that band-like model should be used if the inter-molecular
  hopping element $t$ is larger or comparable to the molecule reorganization
  energy $\lambda$.\cite{Coropceanu2007,Kobayashi2013} Following the method in
  Ref. \citenum{Kobayashi2013}, we obtained $\lambda$ of $\sim 0.18$ eV,
  comparable to $t$. Thus, we use the band model to study transport along $y$
  direction. This is also supported by the fact that hopping model predicts a
  mobility that is one-order-of-magnitude smaller than the experimental value
  in a similar molecular crystal.\cite{Tan2009}}

\begin{figure}
\includegraphics[scale=1.1]{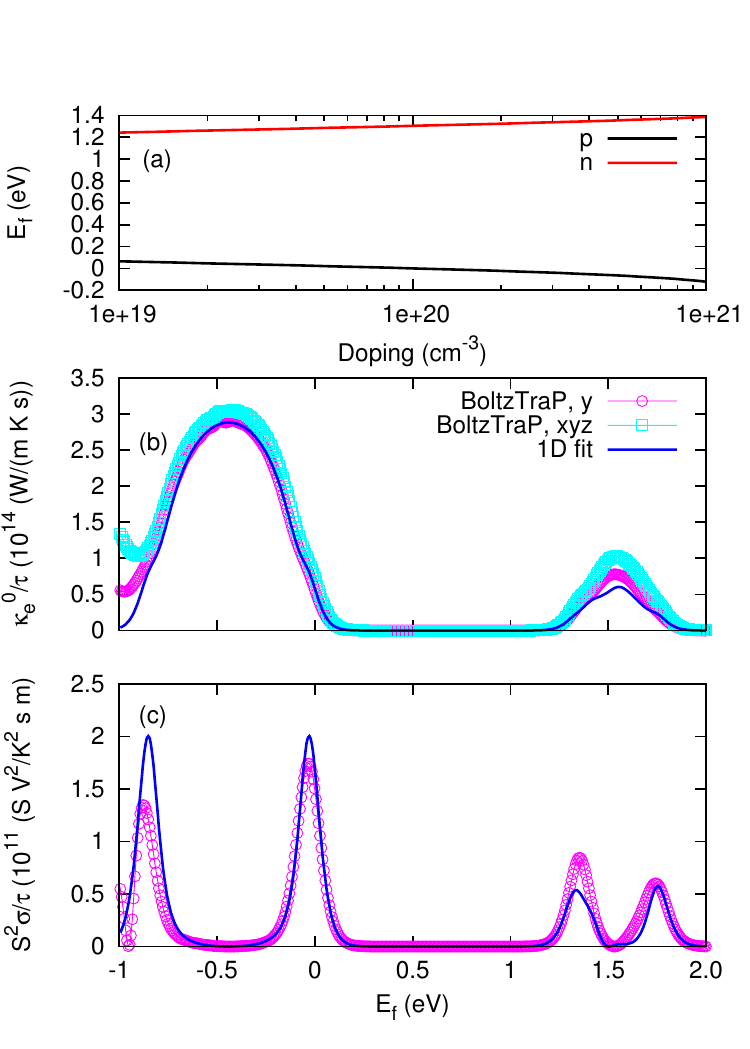}
\caption{\textbf{DFT plus BoltzTraP calculation and model fitting.} (a) The
Fermi level ($E_f$) as a function of electron (red) and hole (black) doping at
$T=300$ K. The weak dependence of the $E_f$ on doping across three decades is
characteristic of quasi-one-dimensional energy band. (b)-(c) The dependence of
$\kappa_e^0/\tau=(\kappa_e+S^2\sigma T)/\tau$ and $S^2 \sigma/\tau$ on $E_f$.
The points are results from DFT plus BoltzTraP calculation, while the lines are
from the 1D tight-binding model with the hopping element $t$ fitted from the
band structure in Fig.~\ref{fig:stru}(b). In (b), $xyz$ means summing
up all three directions, while $y$ means contribution from $y$ direction only. As
can be seen, the contributions from $x$ and $z$ are negligible.}
\label{fig:doping}
\end{figure}

Based on the band structure obtained from
VASP\cite{kresse_efficiency_1996}, we use the
BoltzTraP code \cite{madsen_boltztrap._2006} to calculate
its contribution to the transport coefficients, e.g., 
$(\kappa_e+S^2\sigma T)/\tau$, $S^2\sigma/\tau$, with $\tau$ the
constant relaxation time. Rigid band approximation is used to relate the Fermi
level ($E_f$) position with the electron/hole doping concentration
[Fig.~\ref{fig:doping}(a)]. The weak dependence of $E_f$ on doping within three
decades is due to the van-Hove-like DOS at the band edges. Thus, similar to Q1D
structure, we expect a high Seebeck coefficient shows up in the bulk molecular
crystal. The DFT result is further compared with a 1D model calculation using
the parameters obtained from fitting the band structure (Details in
SI,  I(A)). We find excellent agreement between the
two approaches near the band edges [Fig.~\ref{fig:doping}(b-c)].
This agreement validates the 1D model, which we will use in the
following.

\subsection{Electrical transport.}
\begin{center}
\begin{figure}
\includegraphics[scale=1.4]{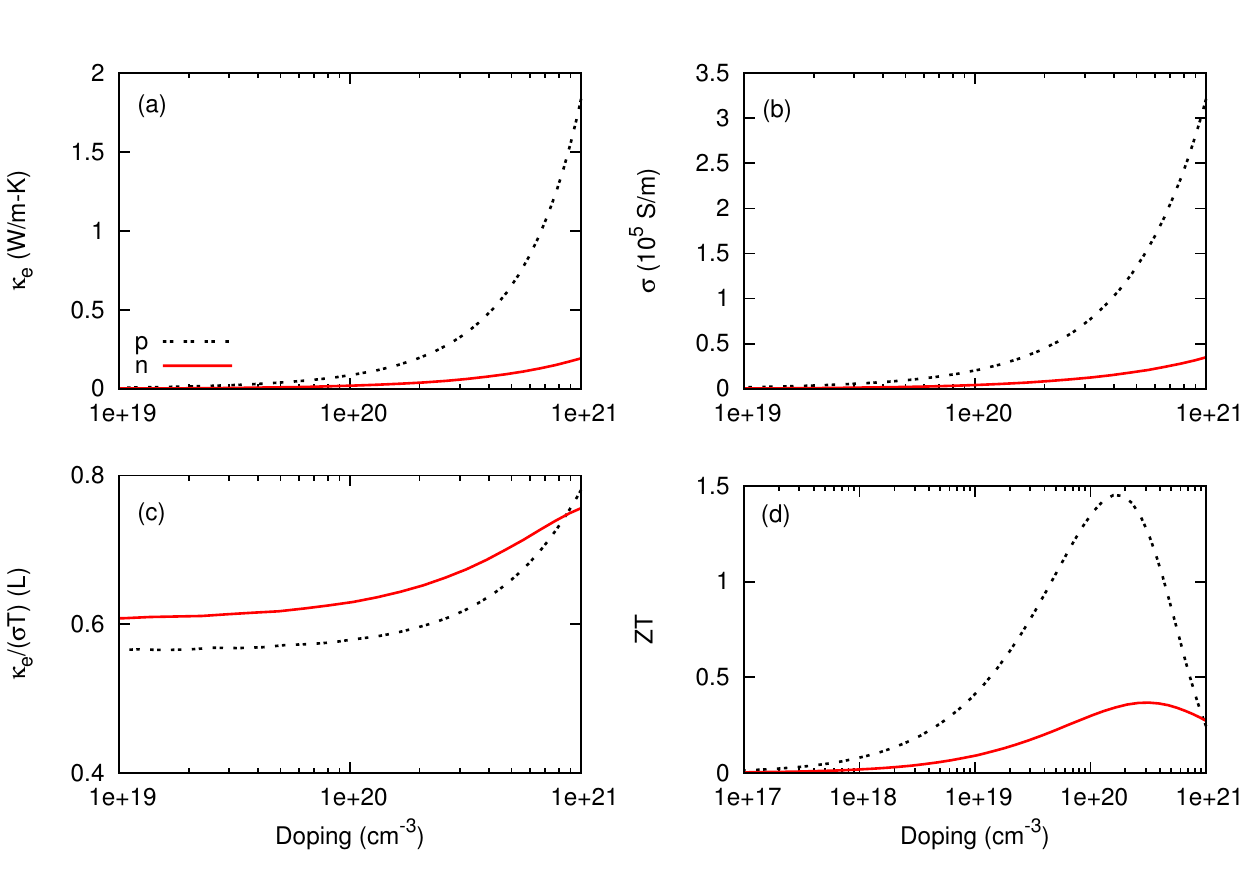}
\caption{\textbf{Thermoelectric transport coefficients at 300 K.} (a)-(b) The electron thermal conductivity $\kappa_e$ and electrical
conductivity $\sigma$ as a function of electron (n) and hole (p) doping. (c) $\kappa_e/(\sigma T)$ as a function of doping, in unit of the Lorentz number $L= \pi^2k_B^2/(3e^2)$, such that the Wiedemann-Franz law corresponds to $\kappa_e/(\sigma T)=L$.
(d) $ZT$ as a function of doping.}
\label{fig:ssk}
\end{figure}
\end{center}

A common approach to obtain the thermoelectric transport coefficients based on
the DFT plus BoltzTraP calculation is to use a constant $\tau$. It is obtained
either from fitting of the experiments, or from the theoretical estimation.
\revision{To estimate the relaxation time, we need to consider different types of carrier scattering process. Defects, charge traps may scatter the carriers strongly, and even invalidate the band-like transport model used here. But these scatterings are extrinsic, and depend on the quality of the sample. 
The charged impurity scattering due to doping is another source of scattering.
To take it account, we should consider the screening of the charged impurity,
which is a problem that deserves separate study. 
Here, we only take into account the intrinsic scattering mechanism, namely the interaction of electrons with acoustic phonons, which is present independent on the quality of the sample.}
Due to the Q1D electronic structure, using the Bardeen-Shockley deformation potential
theory\cite{bardeen_deformation_1950,beleznay2003}, we go beyond  the constant-$\tau$ approach by
taking into account the $k$ dependence of $\tau(k)$ (SI, II)
\begin{equation}
	\frac{1}{\tau(k)} \approx \frac{k_B T D^2}{\hbar^2 C |v_k|}.
\end{equation}
Here, $k_B$ is the Boltzmann constant, $T$ is the temperature,
$C=\partial^{2}{E}/(L_0\partial{\Delta L}^2)$ is the 1D elastic constant,
with $\Delta L = \delta L/L_0$.  $D=\Delta{E}/\Delta L$ is the deformation
potential constant, with $\Delta E$ the energy shift of the valence or
conduction band edge (SI,  I(B)).
This makes our approach different from the common
BoltzTraP calculation, and much closer to the real situation.
The relevant transport coefficients $\kappa_e$, $\sigma$ are shown in
Fig.~\ref{fig:ssk}(a) and (b). We see a better transport performance of the holes. It can be
attributed to their larger hopping  element $t$, leading to larger group
velocity (SI, II). Moreover, a strong deviation of the Wiedemann-Franz law is observed
in Fig.~\ref{fig:ssk}(c), typical at the band edges or low-dimensional structures.
The smaller value of $\kappa_e/(T \sigma) < L$ indicates reduced electron
thermal conductivity, and good thermoelectric performance.

\subsection{Phonon transport.}
\begin{figure}
\includegraphics[scale=0.35]{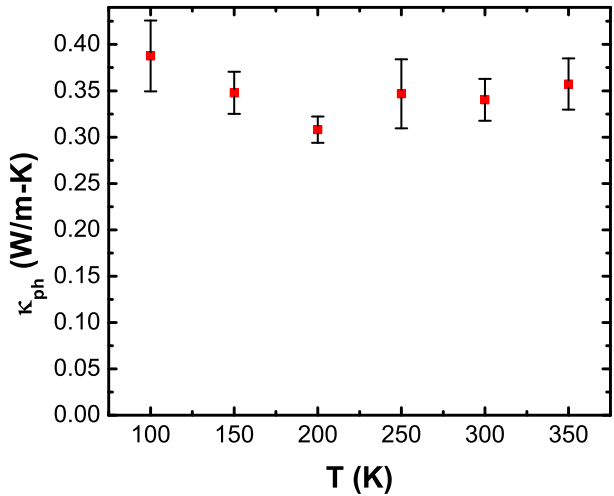}
\caption{\textbf{Phonon thermal conductivity.} The phonon thermal conductivity as a function of temperature obtained from the classical molecular dynamics simulation.}
\label{fig:phkappa}
\end{figure}

To get $ZT$, we also need the phonon thermal conductivity $\kappa_{\rm ph}$,
which was calculated by \revision{equilibrium MD simulation using
  LAMMPS\cite{plimpton_fast_1995}} (Details in SI, III).  
  The following results are obtained from a simulation cell of
  $4\times6\times4$ unit cells in $x$, $y$, and $z$ directions, respectively,
  to overcome the finite size effect (SI,
  III). As shown in Fig.~\ref{fig:phkappa}, the
  thermal conductivity of BDTMC  shows a very weak temperature dependence over
  the temperature range considered ($100$ - $350$ K).  This weak temperature dependence is
  normally observed at high temperature limit, well within the validity of
  classical MD simulation.  We get a value of $0.34 \pm 0.02$ W/m-K
    at $300$ K, which falls into the common range of $0.1 \sim 1$ W/m-K for
    organic molecular crystals.\cite{wang_modeling_2012,chen_2012,zhang_organic_2014}

The low thermal conductivity originates from the weak intermolecular bonding
of VDW nature, in contrast to the strong intramolecular valence
bonding. Different kinds of bonds result in very different frequency/time
scales of inter- and intra-molecular dynamics. The thermal conductivity of the
molecular crystal is dominated by the low-frequency inter-molecular vibrations.
Their frequency mismatch with the intra-molecular vibrations prevents efficient
heat transport between molecules, good for thermoelectric performance. That is,
most of the energy is localized inside each molecule, instead of transferred to
other molecules. This can be seen from the large overshot of $\kappa_{\rm ph}$
as a function \revision{of} correlation time in Fig.~S4.

\subsection{Thermoelectric performance.}
\begin{figure}
\includegraphics[scale=1.1]{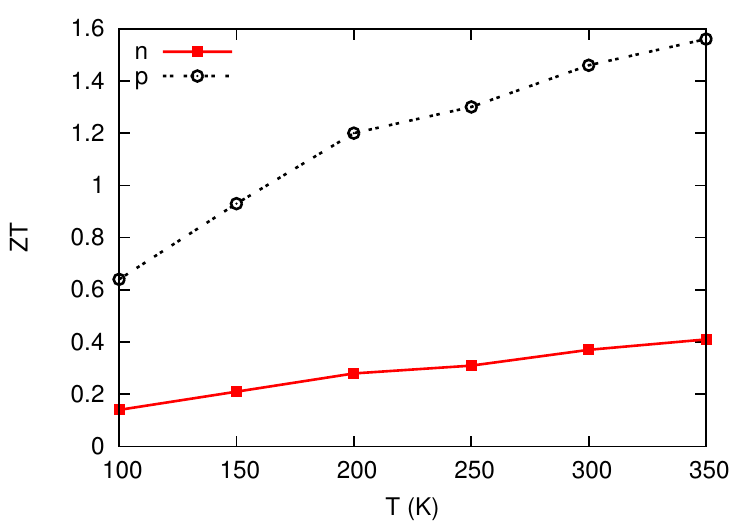}
\caption{\textbf{Temperature dependence of $ZT$ at optimal doping.}}
\label{fig:ztdep}
\end{figure}

We are now in the position of evaluating the thermoelectric figure of merit
$ZT$.  As is shown in Fig~\ref{fig:ssk}(d),  we get an optimal room-temperature
$ZT$ of $0.38$ at electron doping of $\sim 3.2\times 10^{20}$ cm$^{-3}$, and
$1.48$ at hole doping of $\sim 1.57\times 10^{20}$ cm$^{-3}$.  The
corresponding Seebeck coefficients are $-199$ $\mu$V/K for electrons and $266$
$\mu$V/K for holes.  The electron and hole mobility at optimal doping are $\sim
2.4$ and $\sim 12.8$ cm$^2$/V-s, respectively. These results are summarized in
Table~\ref{tab:sum}. 
The following facts are note-worthy: (1) Although doping the organic semiconductors is still challenging in experiments, the optimal doping level we obtained here has been achieved in Pentacene experimentally;\cite{Hayashi_2011} (2) A much higher carrier mobility has been experimentally observed in C8-BTBT;\cite{Yuan_2014} (3) Although the optimal electrical conductivity we obtained here has not been realized so far (the highest experimental value that can be achieved\cite{zhang_organic_2014} is $\sim 11000$ S/m), there is no fundamental difficulty in realizing both (1) and (2) in the same material; (4) On the other hand, a recent study shows that it is possible to reduce the optimal doping by increasing the carrier mobility.\cite{shuai14}
Finally, in Fig.~\ref{fig:ztdep}, we plot the temperature dependence of  $ZT$.  It
shows an approximately linear dependence on $T$. We have checked that $ZT$ saturates at higher $T$ (Fig. S6). Since band-like transport at high $T$ is questionable, we did not show it here.
The $ZT$ value is promising over a wide temperature range ($100$ - $350$ K).

\begin{table*}
  \caption{Summary of important parameters and results for the conduction (CB) and valence band (VB): $t$ the hopping  element in the 1D model, $m^*$ effective mass, $D$ deformation potential, $\mu$ mobility at optimal doping, $S$ Seebeck coefficient at optimal doping ($n$). }	
\begin{tabular}{c |c c c | c c c c c}
\hline
\hline
	& t (eV) & $m^*$ ($m_0$)  & D (eV) & n ($10^{20}$ cm$^{-3}$) & ZT & $\mu$ (cm$^2$/V-s) & S ($\mu$V/K) & $\kappa_e$ (W/m-K) \\
\hline
CB & 0.05 (0.08) & 4.27 & -6.67 & 3.19 & 0.38 & 2.4 & -199 & 0.06\\
\hline
VB & 0.2 & -1.5 & -8.55 & 1.57 & 1.48 & 12.8 & 266 & 0.15\\
\hline
\hline
\end{tabular}
\label{tab:sum}
\end{table*}

The Q1D electrical transport and low phonon thermal conductivity we obtained
here originate from the chemical nature of the inter-molecular bonding. In our
calculation, the conduction and valence band structures of BDTMC are determined
by only one parameter, $t$, reflecting the strength of the $\pi$-$\pi$ VDW
bonding.  By comparing results of the CB and VB, we deduce that stronger
bonding, thus larger $t$, results in better thermoelectric
performance.

Our results, together with recent development in chemical control of the
stacking angle and distance\cite{giri_tuning_2011,dong2013}, point out possible
way of further enhancing the thermoelectric performance by chemical or strain
engineering of $\pi$-$\pi$ overlap integral. DFT calculation confirms this
idea. Within lattice strain of $5\%$, we can achieve more than $20\%$ change of $t$ (Fig.~S2).

\revision{Before closing, we should mention that, as doping is increased, the
impurity scattering rate is also increased proportionally to the dopant
concentration, resulting in a decrease $\sigma$ and $\kappa_e$. But their
ratio should not change much. Considering $ZT=S^2/(\kappa_e/\sigma
T+\kappa_{ph}/\sigma T)$, we expect the $ZT$ to be smaller after including
the impurity scattering. But to include it quantitatively, we need to consider the screening of charged impurity, which
depends on the dimension and dispersion of the electronic band structure.
Since the main idea of this work does not rely on these details, we leave them
to future study.}

\section{Conclusions}
To summarize, we propose a novel approach to search for highly-efficient
thermoelectric materials.  That is to explore or engineer low-dimensional
electronic structure in phonon-glass bulk crystals. Through atomistic
simulation, we have shown that the $\pi$-$\pi$ stacking molecular crystals are
particularly suitable for such an approach. Both low phonon thermal
conductivity and 1D electronic structure originate from the $\pi$-$\pi$
bonding. Besides proving the principle, our results also show the promising
potential of $\pi$-$\pi$ stacking organic crystals as efficient thermoelectric
material. Although we only considered organic crystals in this work, the idea
should be equally applicable to inorganic materials.  Indeed, during the preparation of
this work we became aware of a recent work, exploring similar idea in inorganic
compounds\cite{Bilc2015}.

\section*{Acknowledgements}
J.-T.L. acknowledges stimulate discussions with members of the
Danish-Chinese center for molecular nanoelectronics, from which the idea was
initialized. J.-T.L.  was supported by the National Natural Science Foundation
of China (Grant No. 11304107 and 61371015).  N.Y. was supported by the National
Natural Science Foundation of China (Grant No.11204216), and the
Self-Innovation Foundation of HUST (Grant No.2014TS115).  The authors thank the
National Supercomputing Center in Tianjin (NSCC-TJ) and Shanghai, the High
Performance Computing Center experimental testbed in SCTS/CGCL for providing
help in computations.

%\bibliography{te,thermoelectric}
\providecommand{\latin}[1]{#1}
\providecommand*\mcitethebibliography{\thebibliography}
\csname @ifundefined\endcsname{endmcitethebibliography}
  {\let\endmcitethebibliography\endthebibliography}{}

\end{document}